# Airborne Geophysical Surveys in the North-Central Region of Goias (Brazil): Implications for Radiometric Characterization of Subtropical Soils

Guimarães S.N.P., Hamza V.M. and Justo J. S.

National Observatory (ON/MCT), 20921-400, RJ, Brazil.

**Abstract**

In this work we present progress obtained in analysis airborne geophysical survey data for the north-central region of the state of Goias (Brazil). The data base includes results of both gamma spectrometric and aeromagnetic surveys. Analysis of radiometric data has allowed determination of relative abundances of natural radioactive elements (Uranium, Thorium and Potassium) in the main soil types. The results obtained indicate that most of the subtropical soil types are characterized by Uranium contents of greater than one parts per million (ppm). Only ultisol and oxisol soils are found to have Uranium contents lower than one ppm. Thorium and Potassium abundances also display trends similar to those of Uranium. The K/U ratios fall in the expected range of values for common soils while the Th/U ratios are higher than normal. This latter observation may indicate a characteristic feature of subtropical soils. Alternatively it may be considered as indicative of disequilibrium conditions in radioactive series and consequent underestimation of Uranium in soil layers of the study area. A detailed examination of the radiometric data reveals that the currently adopted system for classification of soils is rather insensitive to variations in radioelement abundances. In this context we point out the possibility of using results of radiometric surveys as a convenient complementary tool in identifying geochemical zoning of soils in subtropical environments. Analysis of aeromagnetic data reveals the presence of a large number of magnetic lineaments, indicative of distinct structural features in subsoil layers. The analytic signal values point to the existence of substantial small scale variations related to lithologic changes. There are indications that deposition of soil types are controlled to a large extent by the system of northeast – southwest trending faults and fracture systems. According to results of radiogenic heat production calculations the cambisol soil in this region are found to have a mean heat production of $3.32 \pm 5.9$ $\mu Wm^{-3}$ while that of ultisol soil is only $0.36 \pm 0.3$ $\mu Wm^{-3}$. The mean heat production of soil layers at the surface is $0.68 \pm 0.4$ $\mu Wm^{-3}$. Heat production values of basement rocks are estimated to be more than $1.3$ $\mu Wm^{-3}$, after corrections for density changes and non-equilibrium conditions of Uranium series.

# 1. Introduction

The intensities of nuclear radiations in any particular terrestrial environment are in general related to the abundances of natural radioactive elements in rocks and soils of that locality (Adams, 1961; Roser and Cullen, 1964; Paul et al, 1982; Wilford et al., 1997). Gamma ray spectrometric methods have been widely used in measurements of radioactive minerals in soils and basement rocks (see for example: Adams and Gasparini, 1970; Iyengar et al, 1980; Mohanty et al, 2004). It has also been used for mapping area extent of geologic rock formations associated with such radioactive elements. Examples of such studies carried out in Brazil include the works of Blum et al. (2003), Sapucaia et al. (2005).

Measurements of gamma radiation at ground level provide information on the abundances of radioactive elements in the top soil layers with thicknesses of less than 40cm. Nevertheless results of airborne measurements are considered representative of radiometric characteristics of basement rocks beneath the soil cover. The reasoning is that soil covers are in general derived from weathering of basement rocks and in addition some degrees of mixing take place, over long geologic time periods, between soil and subsoil layers. Also, soil layers play key roles in several of the geosphere – biosphere interactions. Hence results of gamma spectrometric methods have been widely used as supplementary tools in geologic mapping. Examples of such studies carried out in Brazil include Vasconcelos et al. (1994) and Maas et al. (2003).

Nevertheless very few studies has been carried out for relating results of air borne radiometric surveys with soil types in tropical regions. With the advent of computer processing facilities it has been possible to analyze results of large scale air-borne radiometric data. In the present work we report progress obtained in analysis of airborne gamma spectrometric data carried out in the state of Goias, an important sub tropical region in the central parts of Brazil. The purpose is to map the radiometric characteristics of subtropical soils and its relevance for estimate radiogenic heat production of subsoil layers on regional scales.

## 2. Geology and Soil Cover in the Study Area

According to earlier geologic studies (see for example, Almeida, 1968; Winge, 1984) the structural province of Tocantins is situated in the region affected by cratonic collision processes during the Brasiliano orogenic cycle. The main structural units are the granitic – gneissic – granulitic blocks (Median Massif of Goias) of Achaean age, meta sedimentary belts (Araguaia, Uruaçu and Brasília) of early Proterozoic age, the magma belt of Goias of mid Proterozoic age and the volcano sedimentary sequences (believed to be associated with island arc regions) of Neoproterozoic age. A simplified geologic map of the area is given in Figure (1).

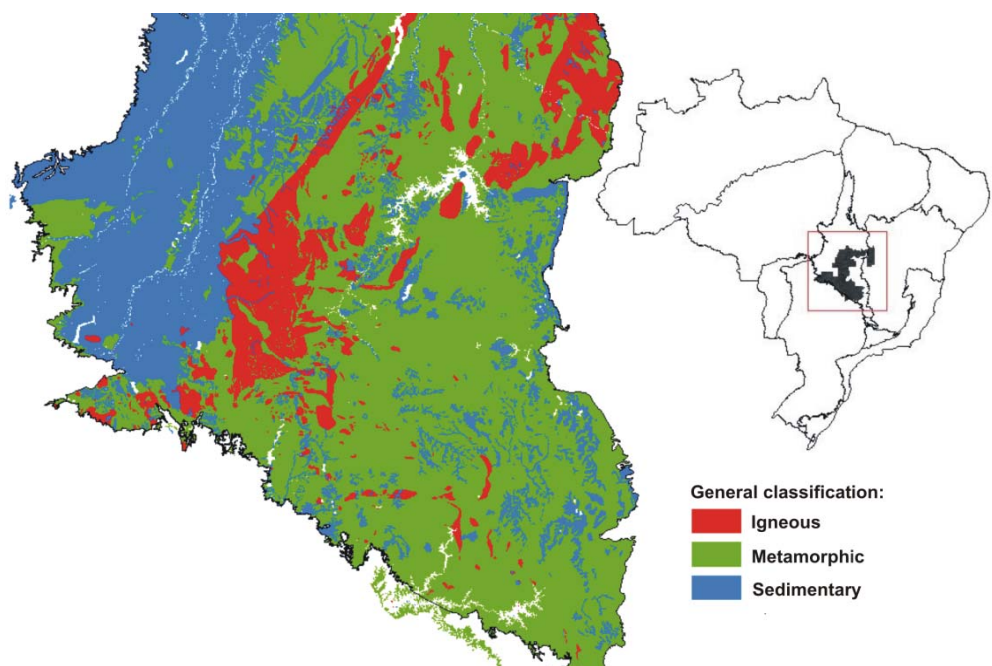

Figure (1) Simplified geologic map of the study area.

The surface features and topography of the study area has also been mapped under the RADAM project of the 1980, by the Ministry of Mines and Energy. A number of studies have been carried out for determining the physical and chemical characteristics of soil cover in the study area (EMBRAPA, 2009). More recently, Kerr (2001) has carried out a more detailed analysis of the results of previous studies. Reproduced in the map of Figure (2) are the eight soil types identified in the study area. Most predominant among these in the study area with large spatial coverage are the oxisol and ultisol soils.

According to Kiehl (1979) the dominant constituents of soils in the study area are quartz, feldspar and colloidal aluminum silicates. Even though individual mineral densities have mean values of about $2,65 \times 10^3$ kg/m$^3$ the bulk densities of most of the soils vary between 1.1 and 1.6 ($\times 10^3$ kg/m$^3$). This is a consequence of the relatively high porosities which vary between 40 a 60%. The mean value of bulk density is $1,5 \times 10^3$ kg/m$^3$.

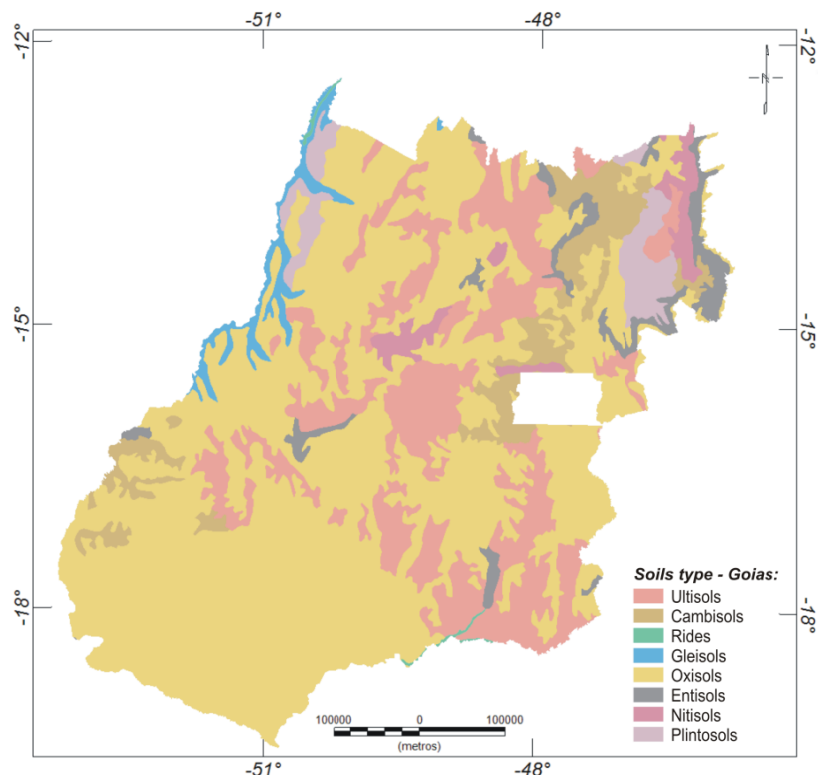

Figure (2) Map of Soil types in the state of Goias.

## 3. Database

The data sets used in the present work has been acquired as part of airborne geophysical surveys of the state of Goias. The data acquisition was carried out during the period of 2004 to 2005 and released since 2010 for academic research by Secretariat for Geology and Mining of the state of Goias (SGM-GO). The locations and area extent covered by the surveys are indicated in the map of Figure (3). The survey work was carried out in five stages, covering regions described as Arenópolis arc (including the Juscelândia Sequence), Mara Rosa Magmatic Arc, western part of Mara Rosa Arc, South Brasilia Belt and north-eastern parts of Goias.

The data sets acquired include measurements of the intensity of gamma radiation emitted by the radioactive elements of Uranium, Thorium and Potassium as well as the characteristics of the local geomagnetic field. These data sets are recorded in separate channels, along with information on flight altitude and atmospheric conditions. The flight lines were set in north − south direction, has spacing of 500m and altitude of 100m. Quality control tests were carried out not only prior to data acquisition, but also during and after the surveys.

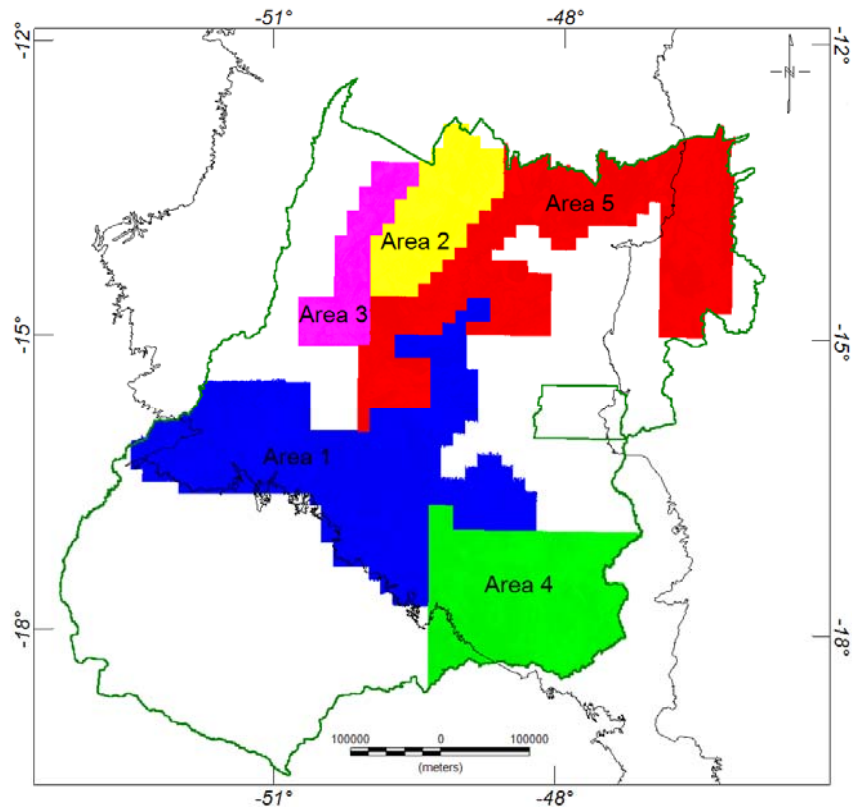
Figure (3) Areas covered in airborne geophysical data acquisitions in the state of Goias.

## 4. Methodology Used in Data Processing

Most of the data processing in the present work has been carried out using the computational package Geosoft®, Oasis Montaj. Initially the raw data were corrected for the perturbing effects of technical survey operations (LAG and Heading effects and altitude variations). In addition, procedures were adopted for filtering, leveling and micro leveling operations, as per standard data processing techniques (see for example, Hood and Ward, 1969; Guimarães and Hamza, 2009). Following this stage a suitable grid system was set up and homogenized data sets derived from the raw data sets for the chosen grid system, using suitable interpolations schemes. The grid size used for interpolations is 125m, which is in accordance with Nyquist criteria. The procedures employed included the method of Minimum Curvature – RANGRID for the geomagnetic data sets and the technique of bi-cubic splines – BIGRID for radiometric data sets. The unification of data sets was achieved using the standard techniques of suture (Geosoft®). The aeromagnetic data were further corrected for the effects of diurnal variations and of the reference magnetic field.

The values of primary radiometric data of air borne surveys are in units of counts per second (cps). These were transformed into values of relative abundances of U, Th and K using conversion factors specific to instrumentation and sensor systems employed and characteristics of each survey operation. In general, these factors depend on the sensitivity and geometry of detectors used and the survey altitude. Such factors are known as sensitivity coefficients (IAEA, 1991) and are specific for the measuring equipment used in each data collection. The values of sensitivity coefficients relevant for data employed in this work are listed in Table 1.

Table (1) – Sensitivity coefficients used in this work.

| Area | Equipment | Sensitivity Coefficients | | |
|---|---|---|---|---|
| | | K (cps to %) | U (cps to ppm) | Th (cps to ppm) |
| 1 | PT-FZN | 79,01 | 9,71 | 5,05 |
| | PT-WOT Upto flight 26 | 77,11 | 11,23 | 5,21 |
| | PT-WOT After flight 26 | 78,05 | 13,05 | 5,10 |
| 2 | PR-FAS | 72,10 | 12,87 | 4,41 |
| | PT-WQT | 76,25 | 11,34 | 4,45 |
| 3 | PT-FZN | 78,76 | 8,89 | 4,94 |
| 4 | PT-FZN | 78,76 | 8,89 | 4,94 |
| | PT-WOT | 78,05 | 13,05 | 5,10 |
| 5 | PT-FZN | 80,28 | 12,59 | 4,99 |
| | PT-WOT | 78,50 | 12,15 | 4,57 |

## 5. Results Obtained

*5.1 Airborne Gamma Spectrometry*

In presenting results of the radiometric survey we focus first on large scale regional features present in the study area. The geographic distribution of Uranium and Thorium abundances, illustrated in the maps of Figures (4) and (5), provide the basis for this analysis. Referring to Figure (4) we note that Uranium abundances are in the range of 0 to 2.5 ppm but there are several localized zones of relatively high values (in excess of 1.5 ppm) distributed all over the study area. Though no clear patters are evident at first sight, careful comparative examination of the maps in figures (2) and (4) reveal that lower values of Uranium abundances are roughly related to the presence of ultisol soil and oxisol soil in the study area. Similar patterns can also be seen in the maps of Thorium and Potassium abundances, illustrated respectively in Figures (5) and (6).

The Thorium abundances are in the range of 0 to 20 ppm, while that of Potassium is in the range of 0 to 2.5%. The overall pattern of abundances of radioactive elements Th and K seem to be related to the characteristics of soils types present in near surface layers.

The ternary diagram presented in Figure (7) illustrates the distribution of relative abundances of the three radioactive elements, U, Th and K. The patterns in this figure are quite complex, but it is possible to identify some general trends. Foremost among these are the zones with relative enrichment of Potassium in the central parts of the study area. There are indications of relative enrichment of Uranium along the western border of the study area, while Thorium enrichment occurs along the eastern border.

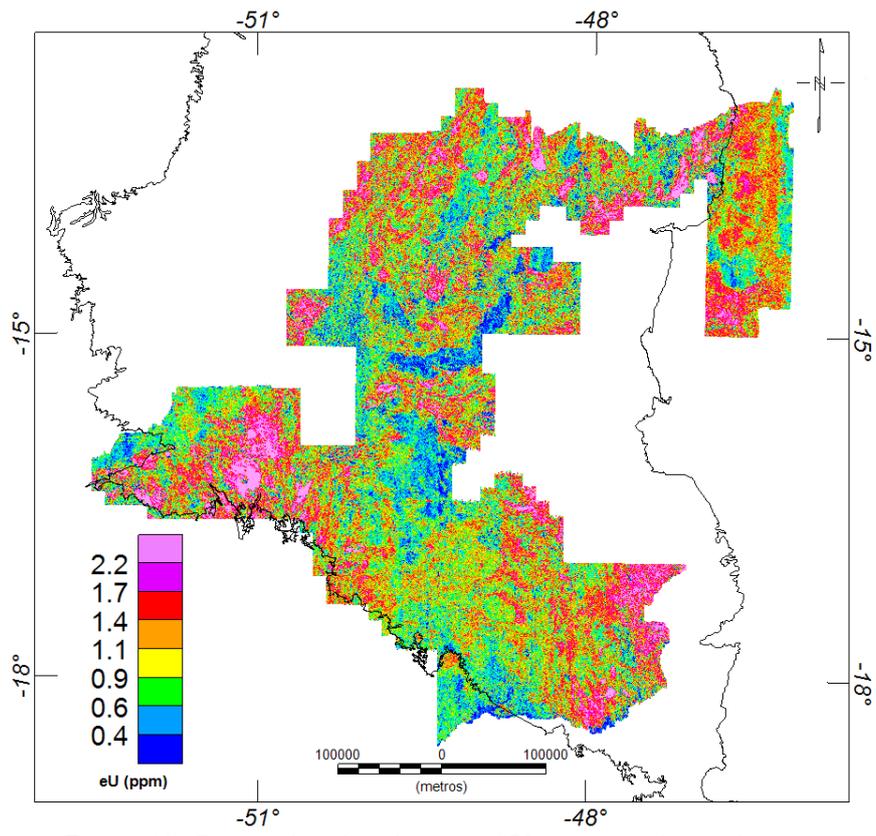
Figure (4) Geographic distribution of Uranium in the study area.

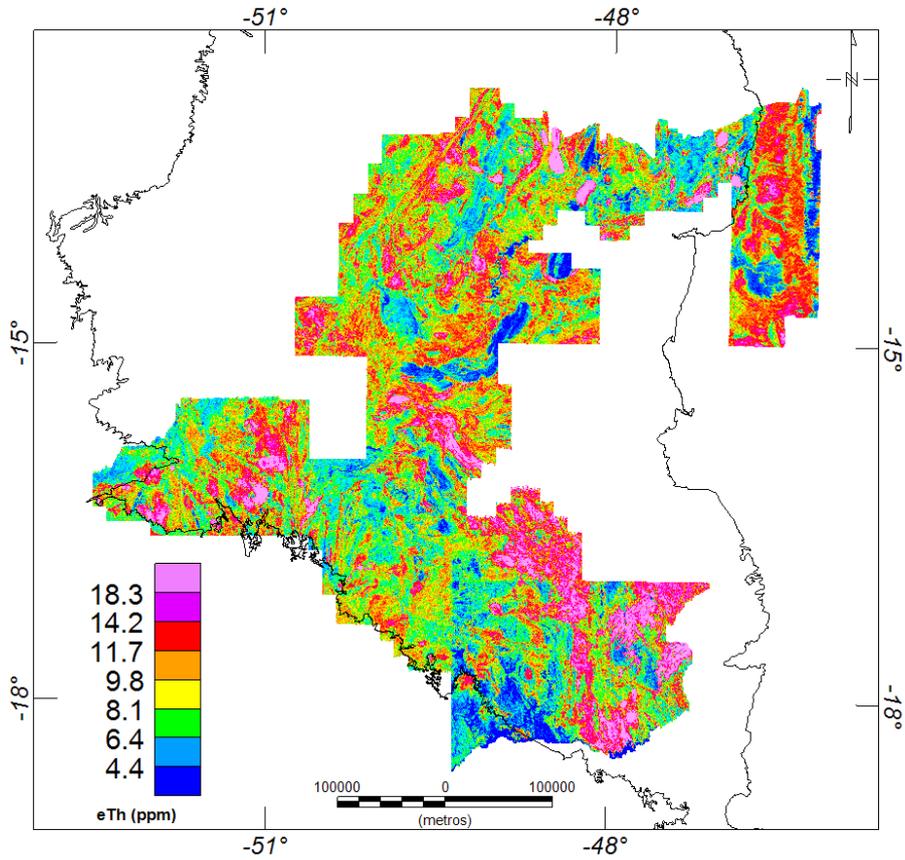
Figure (5) Geographic distribution of Thorium in the study area.

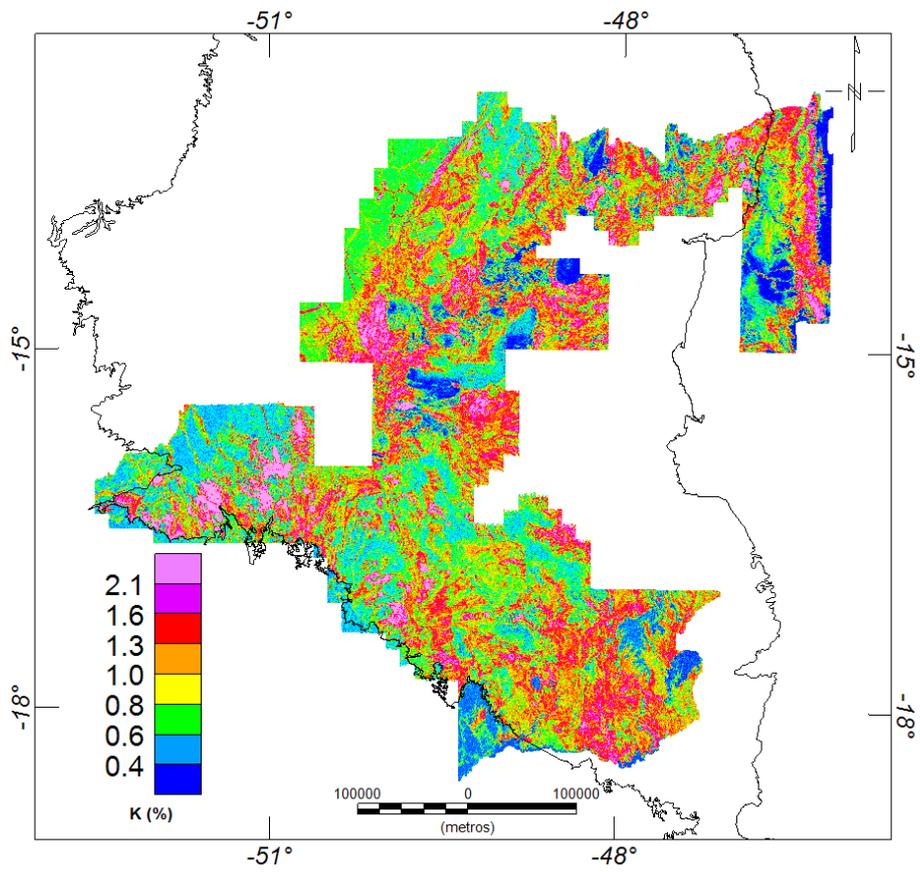

Figure (6) Geographic distribution of Potassium in the study area.

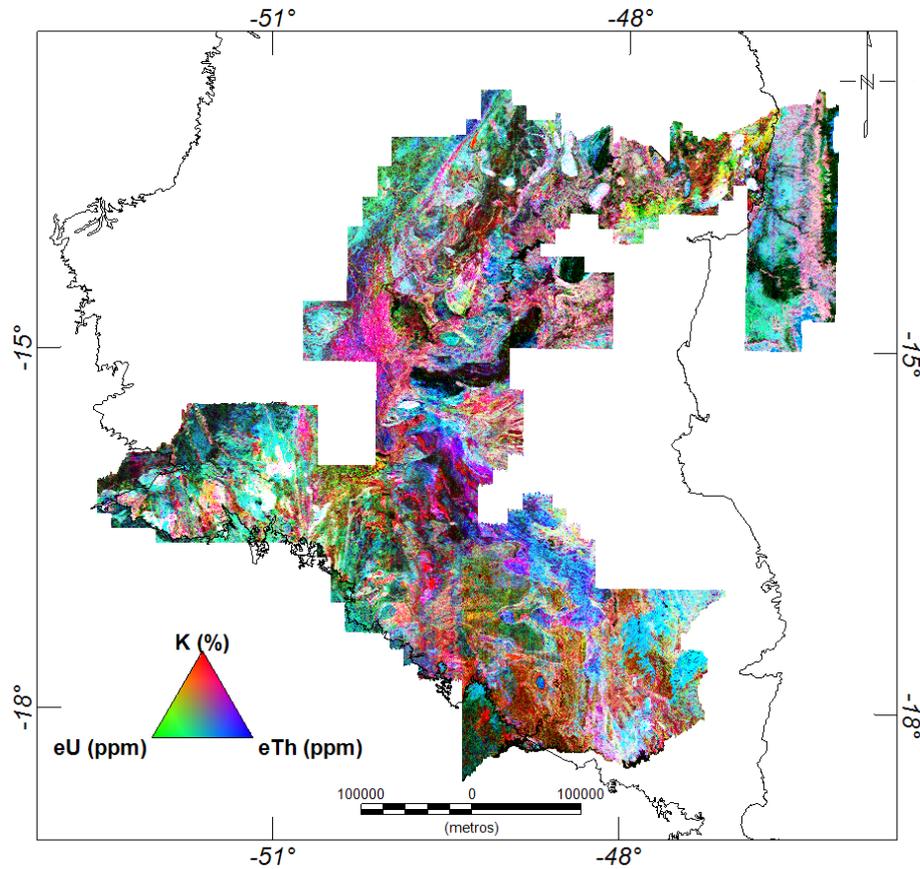

Figure (7) Ternary diagram of the radioactive elements (U, Th and K) in the study area.

In an attempt to examine such relations in detail, grid averaged values of radioactive elements and overall mean values were computed. The mean value of Uranium in the study area is 1.28 ± 1.06 ppm while that of Thorium is 11.4 ± 8.5 ppm. The large values of standard deviations of the estimates obtained are indications of substantial variations in the lithologic characteristics of soil types in near surface layers. These results also reveal that Th/U ratios in the study area are high (> 8). It is possible this is a characteristic feature of radioelement abundances of the study area. The alternate possibility is that disequilibrium conditions of radioactive series exist in soil layers and have contributed to underestimation of Uranium in surface layers. The mean value of Potassium is 1.2 ± 0.8 % which imply that K/U is of the order of $1x10^{4}$, a value typical of common crustal rocks.

*5.2 Aeromagnetic Survey*

In presenting results for the aeromagnetic survey the focus has been on identification of structural features present at shallow depths. In such a context the standard practice is to outline the magnetic lineaments (deduced from the first derivative of the crustal field). The geographic distribution of the main lineaments identified in the study area is illustrated in the map of Figure (8). The outstanding features in this map are the two systems of magnetic lineaments with distinctly different directions. In the southern and north eastern parts of the study area the orientations of the magnetic lineaments are predominantly southeast to northwest, whereas these are southwest to northeast in the northern and western parts.

According to results of geologic studies the preferred directions of structural features of Precambrian age in the study area is predominantly south west to northeast. Hence we deduce that the presence of lineaments of SE – NW direction in the southern parts is a product of tectonic activities of relatively more recent times.

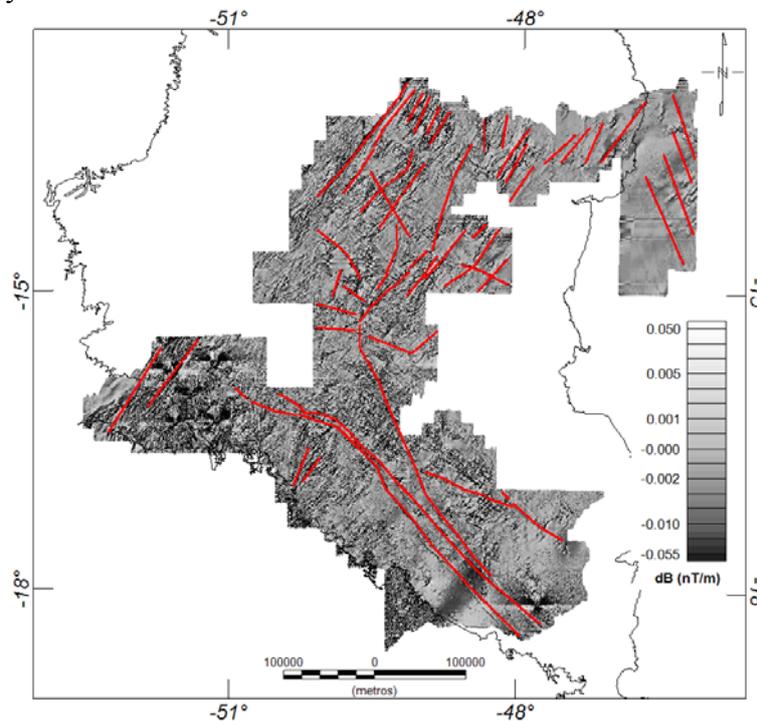

Figure (8) Magnetic lineaments identified in the study area.

The spatial distribution of the analytic signal, which is calculated as the modulus of the second derivative in the three directions of the magnetic field provided additional information on the characteristics of structural features in the study area. The technique is often considered as the best for outlining the borders of subsurface bodies with contrasts in magnetic properties. The map of Figure (9) illustrates the geographic distribution of the analytic signal.

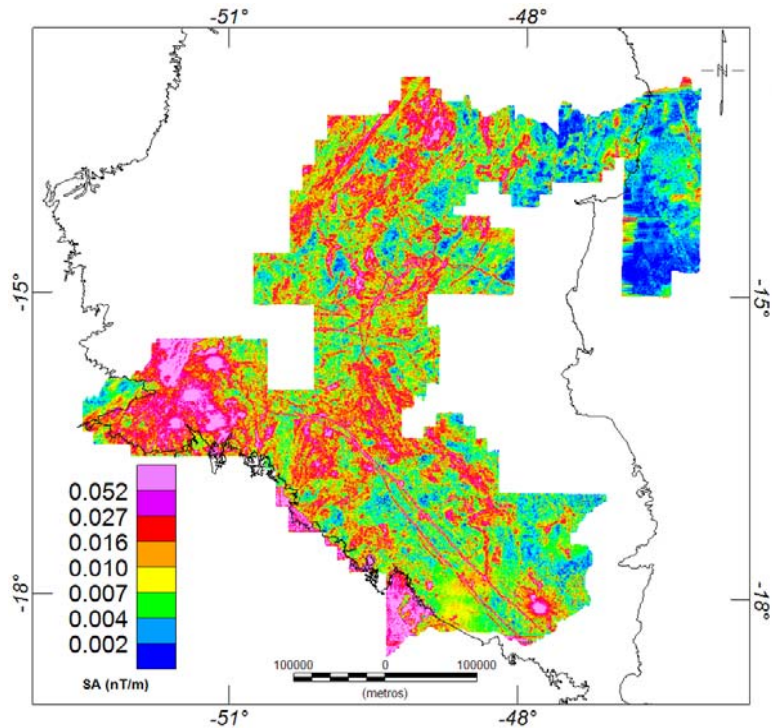

Figure (9) Amplitudes of analytic signal in the study area.

As can be seen in this map the magnitudes of analytic signal are in the range of -0.9 to 45 nT/m with an average value of 0.03 ± 0.2 nT/m. The wide range of values is an indication that contrasts in magnetic properties occur on local scales in the subsurface layers beneath the soil cover in the study area. Note that the amplitude analytic signal is large in the western and central parts compared to that in the north eastern parts.

**6. Discussion**

*6.1 Radiometric Characterization of Soil Types*

In this item we discuss the results relevant to the study of soil types in the study area. In doing so we compare regional scale features present in the maps of abundances of radioactive elements with those of soil types. Superposition of the relevant shape files has allowed determination mean values of U, Th and K in the main soil types. The results obtained are presented in Table (2). Note that plintosol, cambisol, entisol, gleisol and nitisol soils are characterized by relatively high values of Uranium and Thorium, when compared with those of oxisol and ultisol soils. The Th/U ratios are in the range of 5 to 10 for all of the soil types. The Potassium abundances also follow a similar trend, the only exception being low mean value for plintosol soils. The possibility that widespread use of potash rich fertilizers have interfered with the results airborne radiometric surveys cannot entirely be ruled out.

Table (2) – Mean and standard deviation of U, Th and K in soil types of the study area.

| Soil Type | eU (ppm) | eTh (ppm) | K (%) |
|---|---|---|---|
| **Plintosol** | 1.46 ± 0.9 | 10.97 ± 6.8 | 0.70 ± 0.7 |
| **Cambisol** | 1.45 ± 1.4 | 11.90 ± 8.3 | 1.48 ± 1.0 |
| **Entisol** | 1.26 ± 1.0 | 8.94 ± 7.1 | 1.17 ± 0.8 |
| **Gleisol** | 1.23 ± 0.8 | 10.17 ± 4.4 | 0.96 ± 0.6 |
| **Nitisol** | 1.11 ± 0.7 | 11.39 ± 5.9 | 1.24 ± 0.8 |
| **Oxisol** | 0.80 ± 0.5 | 4.81 ± 5.2 | 0.75 ± 0.8 |
| **Ultisol** | 0.63 ± 0.6 | 5.86 ± 5.8 | 0.92 ± 0.5 |

Though the variations in the abundances of radioactive elements in near surface layers is a problem in mapping radiometric characteristics of basement rocks it has potential applications in classification of soil types. For example, the variations in the abundances of Uranium may be used as a tracer and also as a means of identifying sub classes of soils. As an illustrative example we present in Figure (10) Uranium variations in a segment of cambisol soil in the north eastern part of the study area. It is clear that cambisol soil in this locality is characterized by large scale variations in the radioelement abundances. Even though the overall mean value of Uranium in cambisol soil is 1.4 ± 1.3 ppm it is fairly simple to note the presence of several localities where abundances are greater than 2.5 ppm.

Such variations are indicative of rapid changes in compositional structure of the soil over relatively short distances. It is clear that geochemical zoning associated with radioelement abundances are not readily identifiable by standard procedures in soil classification. We conclude that results of airborne radiometric surveys can be used as a convenient complementary tool in understanding the fine structure of geochemical zoning in subtropical soils.

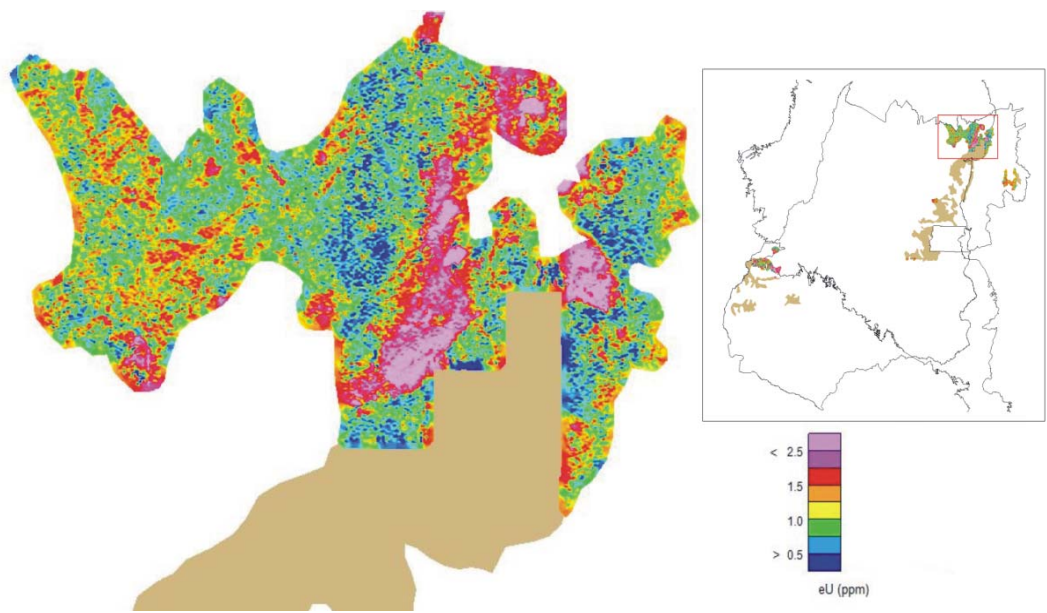

Figure (10) Variations in the U contents of Cambisoil.

*6.2. Radiogenic Heat*

Abundances of Uranium, Thorium and Potassium calculated from results of airborne radiometric surveys have also allowed determination of radiogenic heat production values for the study area. The procedures adopted are derived from the work of Hamza and Beck (1972) and Hamza (1973). The relevant equation used for this purpose is:

$$H = 10^{-11}\left(9.51 C_U + 2.56 C_{Th} + 3.5 C_K\right) \qquad (1)$$

where H is the rate of heat production in units of W/kg, and $C_U$, $C_{Th}$ and $C_K$ are the concentrations of the radioactive elements. It is common practice to express concentrations of Uranium and Thorium in units of parts per million (ppm) and that of Potassium in units of percent (%). The numbers in equation (1) are conversion factors derived from radioactive decay schemes of individual isotopes, described in the earlier work of Hamza (1973). The volumetric heat production values (A) were calculated as the product of heat production per unit mass (H) and the density (ρ) of source material. For mean soil density of $1.5 \times 10^{-3}$ kg/m$^3$, the relation volumetric heat production, in units of μW/m$^3$, is:

$$A = 0.1428\,C_U + 0.0383\,C_{Th} + 0.0522\,C_K \qquad (2)$$

In using equation (2) it has been assumed that local soil types in general have a density of 1.5 kg/m$^3$. The heat production values calculated for the main soil types and basement rocks are given in Table (3).

Table (3) Heat production values of the main soil types in the study area.

| Soil Type | Heat Production (μW m$^{-3}$) |
|---|---|
| Cambisol | 3.32 ± 5.9 |
| Plintosol | 0.66 ± 0.4 |
| Nitisol | 0.66 ± 0.3 |
| Gleisol | 0.61 ± 0.3 |
| Entisol | 0.58 ± 0.4 |
| Ultisol | 0.36 ± 0.3 |
| Oxisol | 0.34 ± 0.9 |

Most of the soil types have heat production values of less than 0.7μW/m$^3$. The only exception is cambisoils which have a relatively high value of 3.32 μW/m$^3$. The weighted mean value of heat production of soils the study area is 0.68 ± 0.4 μW/m$^3$. This is in reasonably good agreement with the value obtained by Roque and Ribeiro (1997) for core samples of carbonate sequences in shallow boreholes. However these are relatively less than the mean values calculated for a large number of basement rock samples collected from outcrops in the study area by Iyer et al (1984) and also values inferred from crustal seismic velocities by Alexandrino and Hamza (2008) and Hamza et al. (2010). The geographic distribution of heat production values obtained is illustrated in the map of Figure (11). There are a number crustal segments in the eastern parts of the study area where heat production is higher than 1μW/m$^3$, the common range of variation being in the interval of 0.2 to 1.2 μW/m$^3$. There are indications that much of the small scale variations are related to variations in lithologic types mapped in geologic studies.

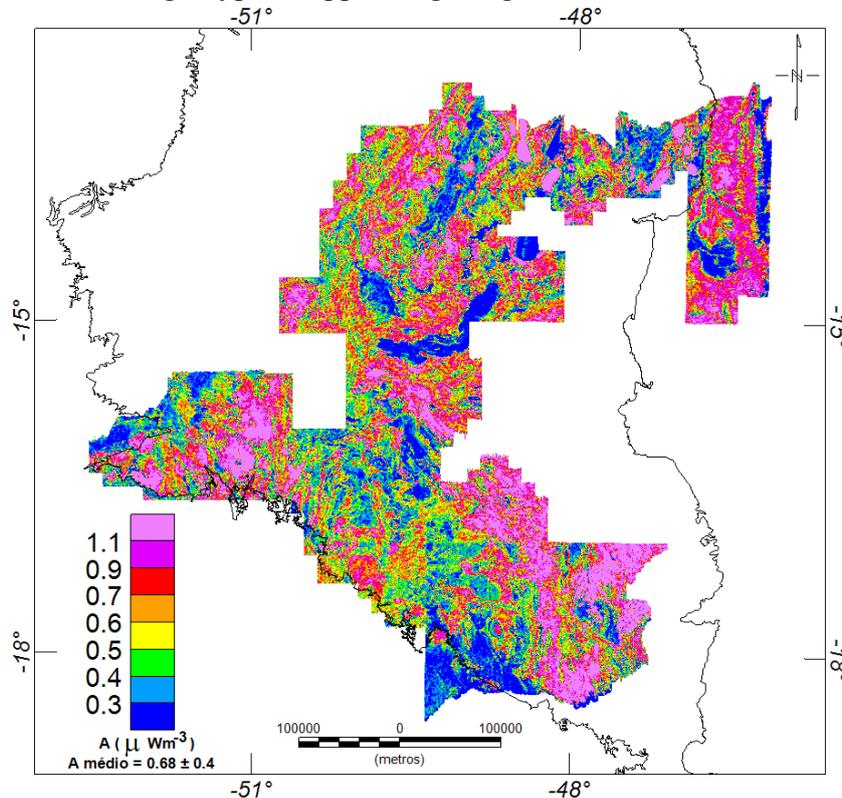

Figure (11) Distribution of radiogenic heat production in the study area.

## 7. Conclusions

In this work progress obtained in analysis airborne geophysical survey data for the north-central region of the state of Goiás (Brazil) has been examined. Analysis of aeromagnetic data has revealed the presence of two distinct sets of magnetic lineaments in the study area: one associated with post metamorphic consolidation of basement rocks during Precambrian times and a second one associated with tectonic activities of later times. The analytic signal values of the magnetic field point to the existence of substantial small scale variations related to lithologic changes.

The results of radiometric survey indicate that most of the common soil types are characterized by Uranium contents of greater than one ppm. Only ultisol soil and oxisol soil are found to have Uranium contents lower than one ppm. Thorium abundances are in the range of 5 to 30 ppm and have trends similar to those of Uranium. However, the Th/U ratios are higher than normal, indicating a specific radiometric characteristic of the study area. Nevertheless the possibility of such high ratios arising from non-equilibrium conditions of the radioactive series in soft sedimentary layers cannot entirely be ruled out. Another important result of the present study has been identification of sharp changes in radioelement concentrations within individual soil types. Results of air-borne radiometric surveys may be used as a convenient complementary tool in identifying such variations.

The results have also allowed determination of radiogenic heat of the main soil types. Thus cambisol soil in this region is found to have a mean heat production of $3.32 \pm 5.9$ $\mu Wm^{-3}$ while that for ultisol soil is $0.36 \pm 0.3$ $\mu Wm^{-3}$. The mean heat production of soil layers at the surface is $0.68 \pm 0.4$ $\mu Wm^{-3}$.

## 8. Acknowledgments:

We thank Companhia Pesquisa de Recursos Minerais - CPRM, Secretaria de Geologia Mineração - SGM-GO, Empresa Brasileira de Agricultura e Pecuária - EMBRAPA and Observatório Nacional - ON for providing data sets used in this work. The first author of this work is recipient of a scholarship granted by Coordenadoria de Apoio á Pesquisa e Ensino Superior - CAPES.